\documentclass{elsart}               

\usepackage{graphicx}          
                               \usepackage[utf8]{inputenc}

\usepackage{amsmath}
\newenvironment{proof}{\par\noindent\textbf{Proof.}\ }{\hfill$\square$\par}
\usepackage[version=4]{mhchem}
\usepackage{siunitx}
\usepackage{longtable,tabularx}
\newtheorem{lemma}{Lemma}[section] 
\usepackage{cite}
\usepackage{amsmath,amsfonts,derivative}
\usepackage{subcaption}
\usepackage{algorithm}
\usepackage{algorithmic}

\usepackage{textcomp}
\usepackage{xcolor}
\usepackage{comment}
\def\BibTeX{{\rm B\kern-.05em{\sc i\kern-.025em b}\kern-.08em
    T\kern-.1667em\lower.7ex\hbox{E}\kern-.125emX}}

\usepackage{pgfplots}
\pgfplotsset{compat=1.18}

\usepackage{graphicx} 
\usepackage{tikz}
\usetikzlibrary{calc, patterns, decorations.pathmorphing,decorations.markings}
\usepackage{algorithm}

\pgfplotsset{compat=newest}
\usetikzlibrary{external}
\usepgfplotslibrary{external}

\usepackage{diagbox} 
\newtheorem{theorem}{Theorem}

\begin{document}

\begin{frontmatter}

\title{ On the embedding transformation for optimal control of multi-mode switched systems} 

\thanks[footnoteinfo]{This research is supporting, in part, by the Office of Naval Research
grant N00014-21-1-2481.Any opinions, findings, conclusions, or recom-
mendations detailed in this article are those of the author(s), and do not
necessarily reflect the views of the sponsoring agencies.}

\author[1]{Masoud S. Sakha}\ead{masoud.sakha@ufl.edu} and   
\author[1]{Rushikesh Kamalapurkar}\ead{rkamalapurkar@ufl.edu}               

\address[1]{Department of Mechanical and Aerospace Engineering, University of Florida, Gainesville, FL, USA}  


\begin{abstract}                          
This paper develops an embedding-based approach to solve switched optimal control problems (SOCPs) with an arbitrary number of subsystems. Initially, the discrete switching signal is represented by a set of binary variables, encoding each mode in binary format. An embedded optimal control problem (EOCP) is then formulated by replacing these binary variables with continuous embedded variables that can take intermediate values between zero and one. Although embedding allows SOCPs to be addressed using conventional techniques, the optimal solutions of EOCPs often yield intermediate values for binary variables, which may not be feasible for the original SOCP. 
To address this challenge, a modified EOCP (MEOCP) is introduced by adding a concave auxiliary cost function of appropriate dimensionality to the main cost function. This addition ensures that the optimal solution of the EOCP is bang-bang, and as a result, feasible for the original SOCP.
\end{abstract}

\end{frontmatter}
\section{Introduction}
Over the past few decades, switched systems have gained significant attention for their ability to model a wide range of real-world phenomena, with applications including automotive systems, automation, robotics, and space exploration. \cite{SCC.Shaker.Wisniewski2009,  SCC.Seyed.Sakha.Shaker.ea2018, SCC.Liberzon.Morse1999, SCC.Shorten.Wirth.ea2007, SCC.Lin.Antsaklis2009,  SCC.Leth.Wisniewski2012, SCC.Greene.Sakha.eatoappear, SCC.Xu.Liu.ea2023a, SCC.Yuan.Zhao.ea2024a}.

In automotive systems, for example, the operation of a hybrid electric vehicle can be seen as a switched system, where control alternates between electric and fuel-driven modes to optimize fuel efficiency and performance~\cite{SCC.Singh.Bansal.ea2019}. In industrial automation, on-off relays are commonly used to switch machinery or processes between active and idle states. These relays serve as a discrete control mechanism for regulating operations such as material handling, sorting, and assembly. The system switches between different states to manage various stages of production, contributing to efficiency and precision in automated manufacturing lines~\cite{SCC.Madala.Sagar2018}.
In robotic systems, switched dynamics naturally arise when the controller alternates between different operation modes, such as motion, interaction, or manipulation phases, depending on the task requirements. This switching enables robots to adapt to dynamic environments and efficiently accomplish complex tasks~\cite{SCC.Zhang.Wang.ea2023}.
In space systems, switched systems are fundamental to orbital maneuver problems, such as orbital changes or rendezvous maneuvers with on-off thrusters~\cite{SCC.Heydari.Balakrishnan2013b}. Here, control alternates between thrusting and coasting modes to achieve precise positioning and orientation adjustments, enabling complex orbital maneuvers.

As a subclass of hybrid dynamical systems, switched systems consist of a finite set of subsystems and a logical framework that governs transitions between them. The coexistence of discrete and continuous dynamics gives rise to complex behaviors, particularly when the system transitions between different modes based on specific conditions.

The optimal control problem for switched systems involves determining the optimal control input, including the switching logic, to maximize performance while accounting for discrete transitions between different subsystems (or modes) of the system. Due to the hybrid nature of switched systems, traditional optimal control methods, which rely solely on the gradient of the cost function, are insufficient for solving such problems.~\cite{SCC.Xu.Antsaklis2003, SCC.Zhu.Antsaklis2014, SCC.Zhao.Gan.ea2020}.
 
Switched systems are generally classified into two main categories: those with externally forced switching (EFS), which is the focus of this study, and those with internally forced switching (IFS)~\cite{SCC.Zhu.Antsaklis2014}. Approaches for solving switched optimal control problems (SOCPs) with EFS can be grouped into two primary methods: two-stage optimization and embedding approaches.

Two-stage optimization methods, first introduced in \cite{SCC.Xu.Antsaklis2000}, consist of two layers: the inner layer optimizes switching times, while the outer layer focuses on switching sequences. Many studies in this category concentrate on optimizing switching times within a fixed sequence \cite{SCC.Xu.Antsaklis2000, SCC.Xu.Antsaklis2003, SCC.Xu.Antsaklis2004}. The primary challenge in optimizing switching sequences lies in the combinatorial growth of sequence options with the number of switching events. To address combinatorial explosion, a mode insertion technique was developed in \cite{SCC.Axelsson.Wardi.ea2005} to avoid combinatorial explosion by updating the switching sequence iteratively, inserting a new mode at an optimal time and duration in each interaction. In \cite{SCC.Axelsson.Wardi.ea2007}, a gradient descent method is employed for mode insertion, addressing two main theoretical questions: convergence in Euclidean spaces of increasing dimensions and the existence of convergent algorithms. This idea is further extended in \cite{SCC.Wardi.Egerstedt.ea2014} to allow multiple mode insertions per iteration, along with a new convergence analysis framework based on Polak’s notions of optimality~\cite{SCC.Polak.Wardi1984} for infinite-dimensional optimization problems.

Unlike two-stage optimization, embedding approaches treat the discrete mode sequence variables as continuous decision variables. The concept of embedding switched systems within a broader class of continuous systems was first introduced in \cite{SCC.Bengea.DeCarlo2005}. This work demonstrates how the embedded state space and cost for switched systems with two and three subsystems can be formulated within the embedded framework. A key theoretical result established in \cite{SCC.Bengea.DeCarlo2005} is that the set of switched trajectories is dense in the set of embedded trajectories, ensuring that every embedded solution can be approximated arbitrarily closely by a switched solution. This approximation is realized through high-frequency switching, where the switched system alternates rapidly between modes to reproduce the effect of a convex combination. If a bang–bang solution exists for the embedded optimal control problem (EOCP), it can be directly recognized as a solution to the original switched optimal control problem (SOCP). Otherwise, the density property guarantees that a suboptimal switched solution can be obtained via the chattering lemma. Later, \cite{SCC.Bai.Yang2007} provided a simplified proof of the density property for the two-subsystem case by employing the Lyapunov convexity theorem, thereby avoiding the use of relaxed systems and the chattering lemma.

More recently, \cite{SCC.Chen.Zhang2017} reinterpreted embedding within a weak topology framework, emphasizing that “closeness” of inputs need not mean pointwise similarity. Two switching signals may look very different in time, yet still be close under a weak topology if they generate nearly the same terminal state or cost. This interpretation shows that embedding is not only a convexification device but also a topological relaxation that unifies existing algorithms. By choosing appropriate weak topologies, one can explain the convergence behavior of different embedding methods and guarantee that solutions of the embedded problem correspond to meaningful switched solutions.

The embedding approach offers a significant advantage over two-stage optimization by eliminating the need to assume the number of switching events. Furthermore, it facilitates the use of traditional optimal control techniques, such as Pontryagin’s minimum principle and dynamic programming. However, a solution to the EOCP is not necessarily feasible for the SOCP, as the switching signals are restricted to specific integer values, whereas the EOCP allows real values within predefined intervals. To address this feasibility issue, \cite{SCC.Bengea.DeCarlo2005} adapted the approach by adding a concave auxiliary function to the cost function. This modification ensures that the modified EOCP (MEOCP) has a bang–bang solution, making it feasible for the SOCP.

The extension of embedding beyond the two- or three-subsystem case was carried out for linear systems in \cite{SCC.Wu.Sun.ea2018}, 
which formulated the switched linear quadratic regulator (LQR) problem with an arbitrary number of subsystems by embedding the switching variables on the probability simplex. This development showed that closed-form optimal switching conditions and Riccati-based feedback laws could be obtained for general $M$-mode linear switched systems, but the analysis was restricted to linear dynamics with quadratic costs. Separately, \cite{SCC.Abudia.Harlan.ea2020} introduced an auxiliary-cost modification to enforce bang–bang solutions in embedded formulations. While effective in resolving feasibility issues for nonlinear two-mode systems, this method did not provide a systematic path for general multi-mode settings, since extending such concave penalties to higher dimensions remains challenging. Taken together, existing embedding approaches therefore face two main limitations: they either address nonlinear switched systems restricted to two subsystems, or they allow an arbitrary number of subsystems but only under linear-quadratic assumptions. In this work, we overcome both restrictions by developing a framework that extends embedding methods to switched systems with nonlinear dynamics and general cost structures. We introduce a binary encoding of the discrete mode variables and, crucially, design a concave auxiliary function that operates in arbitrary dimensions to enforce bang–bang solutions. This provides a systematic method for deriving the embedded optimal control problem (EOCP) from the switched optimal control problem (SOCP) and for constructing the modified EOCP (MEOCP), thereby broadening the applicability of embedding methods well beyond the classical linear LQR setting.

The proposed approach is applied to two benchmark problems. The first is the nonlinear three–tank system, where the objective is to regulate the fluid levels using discrete ON/OFF pumps. The second is the orbital rendezvous problem, in which a spacecraft equipped with perpendicular ON/OFF thrusters must achieve rendezvous with a target satellite moving in a circular coplanar orbit. The simulation results in Section \ref{sec:Simulation} demonstrate the effectiveness of the developed method in both scenarios.
\section{Problem Formulation for Multi-Mode Switched Systems}
Consider a switched nonlinear system with $M$ subsystems defined as
\begin{equation}
    \dot{x} = f_{q}(t, x, u)
    \label{Eq_switching function}
\end{equation}
where $t \in [t_0, t_f]$ represents time, $x \in \mathbb{R}^n$ is a state vector, $q: \mathbb{R}_{\geq 0} \rightarrow \mathcal{Q} := \{0, 1, 2, \cdots, M-1\}$ is a piecewise constant switching signal, $u \in \mathbb{R}^m$ is the control input, and $f_i \in \mathcal{C}^1(\mathbb{R}^n)$ for all $i \in \mathcal{Q}$.

A switching schedule $\sigma$ consists of a switching sequence $\{q_k\}_{k=1}^{N}$ with $q_k \in \mathcal{Q}$, and a corresponding set of switching time instances $\{\tau_k\}_{k=0}^{N}$ satisfying $t_0 = \tau_0 < \tau_1 < \cdots < \tau_N = t_f$, where $0 \leq N < \infty$. The signal $q$ is constant on each subinterval and satisfies $q(t) = q_k$ for all $t \in [\tau_{k-1}, \tau_k)$.

A SOCP is formulated as
\begin{equation*}
\begin{aligned}
    \underset{u(\cdot), q(\cdot)}{\min} \quad & J(t_0, x_0, t_f, u(\cdot), q(\cdot)) \\
    \text{subject to} \quad & \text{(i) } (t_0, x_0, t_f, x(t_f)) \in \mathcal{B} \subseteq \mathbb{R}^{2n+2} \\
    & \text{(ii) } q(t) \in \mathcal{Q}, \quad u(t) \in \mathcal{U} \subseteq \mathbb{R}^m, \quad \forall t \in [t_0, t_f],
\end{aligned}
\end{equation*}
where $\mathcal{B}$ and $\mathcal{U}$ are compact sets. The cost function is defined as
\begin{equation}
\begin{aligned}
J(t_0, x_0, t_f, u(\cdot), q(\cdot)) := & \int_{t_0}^{t_f} \ell_{q(t)}(t, x(t), u(t)) \, \mathrm{d}t 
 + K(t_0, x_0, t_f, x(t_f)),
\end{aligned}
\label{Eq_Cost SW}
\end{equation}
and $x(\cdot)$ is a solution of (\ref{Eq_switching function}) under the control signal $u(\cdot)$ and the switching signal $q(\cdot)$, starting from $x(t_0) = x_0$.

The objective is to determine an optimal schedule $\sigma^* = (q^*, \tau^*)$ that satisfies the constraints and minimizes the cost defined in \eqref{Eq_Cost SW}.

\section{Switched Optimal Control Formulation via Binary Encoding}
The operation mode $q$ can be encoded using a set of binary variables as
\begin{equation}
q(t) = (\bar v_{b-1}(t), \ldots , \bar v_1(t), \bar v_0(t))_2 
= \sum_{i=0}^{b-1} 2^i \bar v_i(t),
\end{equation}
where each $\bar v_i : \mathbb{R}_{\ge 0} \to \bar V := \{0,1\}$ is a binary switching variable, and $b$ is the smallest integer such that 
$2^{b-1} < M \le 2^b$ (equivalently, $b = \lceil \log_2(M) \rceil$, where $\lceil \cdot \rceil$ denotes the ceiling operator).  

We define the binary switching vector as
\begin{equation}
\bar v(t) := [\bar v_0(t), \bar v_1(t), \ldots , \bar v_{b-1}(t)]^\top \in \bar V^b .
\end{equation}
Representing the discrete mode $q(t)$ in binary format offers two key advantages. 
First, it provides a systematic way to extend embedding-based approaches to an arbitrary number of subsystems, since the number of binary variables grows as $\lceil \log_2(M) \rceil$ instead of linearly with $M$. 
Second, the binary structure allows invalid mode indices to be identified and penalized explicitly, which facilitates the design of concave auxiliary functions in the modified embedded problem. 
For example, a system with $M=5$ modes requires only $b=3$ binary variables $\bar v_0, \bar v_1, \bar v_2$, and the mode index is computed as $q(t) = 4\bar v_2(t) + 2\bar v_1(t) + \bar v_0(t)$. 
In this case, the binary combination $(\bar v_2,\bar v_1,\bar v_0)=(1,1,0)$ produces $q=6$, which lies outside the valid set $Q=\{0,1,2,3,4\}$ and is therefore excluded through penalty terms in the MEOCP formulation.

\vspace{2mm}
The SOCP reformulated using binary decision variables takes the form
\begin{equation*}
\begin{aligned}
    \underset{u(\cdot),\,\bar{v}(\cdot)}{\min} \quad & J(t_0, x_0, t_f, u(\cdot), \bar{v}(\cdot)) \\
    \text{subject to} \quad 
    & \text{(i) } (t_0, x_0, t_f, x(t_f)) \in \mathcal{B} \subseteq \mathbb{R}^{2n+2}, \\
    & \text{(ii) } u(t) \in \mathcal{U} \subseteq \mathbb{R}^m,~ \forall t \in [t_0, t_f], \\
    & \text{(iii) } \bar{v}(t) \in \left\{ \bar{v} \in \bar{\mathcal{V}}^b ~\middle|~ \sum_{i=0}^{b-1} 2^i \bar{v}_i \in \mathcal{Q} \right\}, ~ \forall t \in [t_0, t_f],
\end{aligned}
\end{equation*}
where $\mathcal{B}$ and $\mathcal{U}$ are compact sets. Note that condition (iii) excludes any binary combinations corresponding to integers equal to or greater than $M$.

The cost function is expressed as
\begin{equation}
\begin{aligned}
J(t_0, x_0, t_f, u(\cdot), \bar{v}(\cdot)) := & \int_{t_0}^{t_f} \ell(t, x(t), u(t), \bar{v}(t))\,\mathrm{d}t + K(t_0, x_0, t_f, x(t_f)),
\end{aligned}
\label{Eq_Cost_SW}
\end{equation}
where the running cost $\ell(\cdot)$ depends on the active mode encoded by $\bar{v}(t)$.
\section{Embedded Optimal Control Formulation}
Most conventional numerical optimization algorithms based on Pontryagin's Minimum Principle are developed for problems with continuous decision variables. To take advantage of such algorithms, the SOCP with binary variables is transformed into a continuous optimal control problem by embedding the binary switching variables into continuous functions.

The embedded system dynamics are expressed as
\begin{equation}
    \dot{x} = f_e(t, x, U)
    \label{Eq_Em function}
\end{equation}
where $U := [u^\top, v^\top]^\top$ is the augmented control vector, with $u := [u_0^\top, \ldots, u_{M-1}^\top]^\top$ and $v := [v_0, v_1, \ldots, v_{b-1}]^\top$. 
Each $u_i: \mathbb{R}_{\geq 0} \rightarrow \mathcal{U}$, $\forall i \in \mathcal{Q}$, is the control input in vector field $f_i$, and each $v_i: \mathbb{R}_{\geq 0} \rightarrow \mathcal{V} := [0,1]$ is a relaxed version of the binary variable $\bar{v}_i$ introduced in the previous section.

The embedded vector field is defined as
\begin{equation}
    f_e(t, x, U) := \sum_{k=0}^{M-1} V_k(v) f_k(t, x, u_k),
    \label{Eq_fe function}
\end{equation}
where $f_k$ denotes the dynamics associated with mode $k$, and $u_k$ is the control applied under mode $k$.

\vspace{1mm}
Each coefficient $V_k(v)$ is computed from the binary expansion of $k$. Let $k = (k_{b-1} \cdots k_1 k_0)_2$ be the binary representation of $k$, where $k_i \in \{0, 1\}$ for $i = 0, \dots, b-1$. Then
\[
V_k(v) := \prod_{i=0}^{b-1} \left[ k_i v_i + (1 - k_i)(1 - v_i) \right].
\]
The embedded optimal control problem (EOCP) is then given by:
\begin{equation*}
\begin{aligned}
    \underset{U(\cdot)}{\min} \quad & J(t_0, x_0, t_f, U(\cdot)) \\
    \text{subject to} \quad 
    & \text{(i) } (t_0, x_0, t_f, x_f) \in \mathcal{B} \subseteq \mathbb{R}^{2n+2}, \\
    & \text{(ii) } u(t) \in \mathcal{U}^M,\quad v(t) \in \mathcal{V}^b,\quad \forall t \in [t_0, t_f],
\end{aligned}
\end{equation*}
where $\mathcal{B}$ and $\mathcal{U}$ are compact sets.

The cost functional is defined as
\begin{equation}
\begin{aligned}
    J(t_0, x_0, t_f, U(\cdot)) := & \int_{t_0}^{t_f} L_e(t, x(t), U(t))\, \mathrm{d}t  + K(t_0, x_0, t_f, x(t_f)),
\end{aligned}
\end{equation}
where the running cost is
\[
L_e(t, x, U) := \sum_{k=0}^{M-1} V_k(v) \ell_k(t, x, u_k).
\]
Here, $\ell_k$ denotes the instantaneous cost associated with subsystem $k$, and $x(\cdot)$ evolves according to the embedded dynamics \eqref{Eq_Em function} under the control $U(\cdot)$ with initial condition $x(t_0) = x_0$.
\section{Modified Embedded Optimal Control Formulation}
An admissible solution for EOCPs may include switching variables that take any value within the interval $[0,1]$, which prevents such a solution from being directly interpreted as a valid solution for the SOCP. Furthermore, even if each $v_i \in \bar{\mathcal{V}}=\{0,1\}$, the decoded index from $v(t)$ may exceed $M - 1$, thus not corresponding to any of the $M$ original subsystems.

However, bang-bang solutions of the EOCP restrict the switching vector $v(t)$ to values in the discrete set $\bar{\mathcal{V}}^b := \{0,1\}^b$, and further, only those that decode to indices in $\mathcal{Q} = \{0,1,\ldots,M-1\}$. Such solutions correspond to valid mode selections and can therefore be interpreted directly as feasible solutions for the SOCP.

In \cite{SCC.Abudia.Harlan.ea2020}, a modified embedded optimal control problem (MEOCP) was developed for systems with two subsystems. The authors showed that by incorporating a carefully chosen auxiliary concave function into the cost, the resulting EOCP admits bang-bang optimal solutions. However, many practical applications involve more than two subsystems, and extending the MEOCP framework to such cases requires a principled design of auxiliary cost terms in higher dimensions. In this work, we develope a systematic formulation for constructing such functions that enforce bang-bang structure in multi-subsystem settings.

\vspace{1mm}
The cost function in the MEOCP is modified as
\begin{equation}
\begin{aligned}
    J(t_0, x_0, t_f, U(\cdot)) := & \int_{t_0}^{t_f} \big\{ L_e(t, x(t), U(t)) + L_M(v(t)) \big\}\, \mathrm{d}t \\
& + K(t_0, x_0, t_f, x(t_f)),
\end{aligned}
\end{equation}
where $L_M(v): \mathcal{V}^b \rightarrow \mathbb{R}$ is an auxiliary concave penalty function designed to guarantee discrete-valued switching decisions that remain within the valid mode set. It satisfies $L_M(v) = 0$ if $v \in \bar{\mathcal{V}}^b$ or corresponds to a mode $k \in \mathcal{Q}$, and $L_M(v) > 0$ if $v \in \mathcal{V}^b \setminus \bar{\mathcal{V}}^b$ or corresponds to a mode $k \notin \mathcal{Q}$.
A key contribution of this work is the design of the auxiliary cost function $L_M(v)$, which promotes valid, discrete-valued switching decisions within the embedded optimal control framework. The function is defined as
\begin{align}
    \label{eq:LM}
    L_M(v) :=\; & \alpha \sum_{i=0}^{b-1} v_i(1 - v_i) + \beta \sum_{k = M}^{2^b - 1} \prod_{i \mid k_i = 1} v_i, 
               \text{where } k = (k_{b-1}, \ldots, k_0)_2, \nonumber
\end{align}
where $\alpha, \beta > 0$ are design parameters. The parameter $\alpha$ discourages fractional values of $v_i$, whereas $\beta$ penalizes combinations that do not correspond to valid mode indices.
\begin{figure}[ht]
    \centering
    \includegraphics[width=0.6\linewidth]{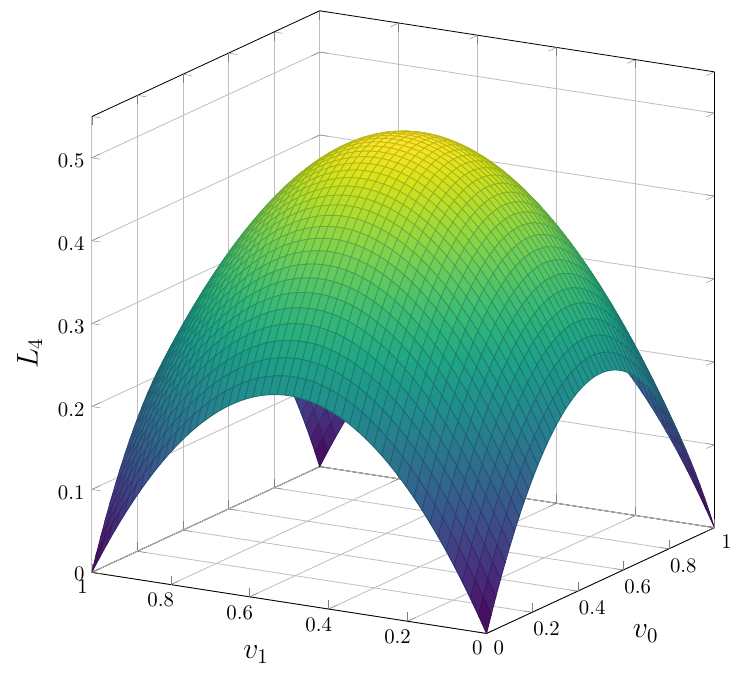}
    \caption{3D Surface Plot of auxiliary cost function for $M=4$ subsystems ($L_4=(v_0-v_0^2)+(v_1-v_1^2)$).}
    \label{Fig:L4}
\end{figure}
\begin{figure}[ht]
    \centering
    \includegraphics[width=0.6\linewidth]{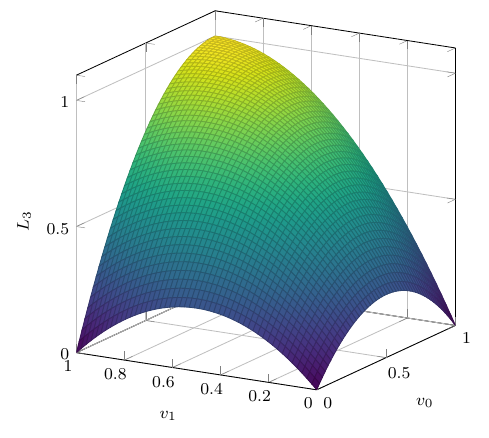}
    \caption{3D Surface Plot of auxiliary cost function for $M=3$ subsystems ($L_3=(v_0-v_0^2)+(v_1-v_1^2)+v_0v_1$).}
    \label{Fig:L3}
\end{figure}

\section{Feasibility of MEOCP Solution via Pontryagin’s Minimum Principle}

To show that every optimal solution of the MEOCP corresponds to a feasible solution of the original SOCP, we apply Pontryagin’s Minimum Principle to characterize the optimal control structure. This principle provides necessary conditions for optimality by introducing an associated costate trajectory and a Hamiltonian function. The Hamiltonian captures the combined effect of the system dynamics, the running cost, and the auxiliary penalty terms, and its pointwise minimization with respect to the control variables characterizes optimal solutions. The Hamiltonian for the MEOCP is defined as
\begin{align}
H(t ,x ,p, U(\cdot))&= \langle~  p~,~ f_e(t,x,U)  ~\rangle  + L(t,x,U)+ L_{M} ( v ),
\end{align}
where $p$ represents the costate and $U := [u^\top, v^\top]^\top$ is the augmented control vector.

The next step is to observe that the Hamiltonian consists of an affine term in $v$ 
and the concave penalty $L_M(v)$ and as a result, is concave in $v$. This is formalized in the following lemma.
\\

\begin{lemma}
\label{lemma:concavity}
If $f:\mathbb{R}^b \to \mathbb{R}$ is an affine function, $g:\mathbb{R}^b \to \mathbb{R}$ a concave function, and  $h:\mathbb{R}^b \to \mathbb{R}$ is defined as $h(v) := f(v) + g(v)$, then $h$ is concave.
\end{lemma}
\vspace{1mm}
\begin{proof}
see  Appendix.
 \end{proof}
 
The concavity property established in Lemma~\ref{lemma:concavity} enables us to prove the main result, which guarantees the feasibility of MEOCP solutions for SOCP.\\

\begin{theorem}
Every optimal solution of the MEOCP is a feasible solution of the SOCP.
\end{theorem}
\begin{proof}
By Pontryagin’s Minimum Principle (PMP) \cite{SCC.Kopp1962}, if $x^*(\cdot)$ 
is an optimal state trajectory and $U^*(\cdot)$ is an optimal augmented control 
 for the MEOCP, then there exists a costate trajectory $p^*(\cdot)$ 
such that the following necessary conditions hold
\begin{align}
H(x^*(t),p^*(t),U^*(t),t) &\leq H(x^*(t),p^*(t),U(t),t), 
&& \forall U(t) \tag{11} \\[6pt]
\dot p^*(t) &= - H_x(x^*(t),p^*(t),U^*(t),t) \tag{12} \\[6pt]
H(t_f) &= 0 \tag{13} \\[6pt]
p(t_f) &= \nabla_{x_f} K(t_0,x_0,t_f,x(t_f)) \tag{14}
\end{align}
To analyze the minimization in (11), define
$U_v(t) := [u^*(t)^\top, v^\top]^\top$
and consider the map
\[
v \mapsto H(t,x,p,U_v,t) = \langle p, f(t,x,U_v,t) \rangle 
+ L(t,x,U_v) + L_M(v).
\]
Here $v = [\,v_1,\dots,v_b\,]^\top$ denotes the embedded mode vector, and $v_i$ is its $i$th component.
The first term, $\langle p, f(t,x,U_v(t))\rangle + L(t,x,U_v(t))$, is 
affine in $v_i$, and the second term, $L_M(v)$, is 
concave in $v_i$. Therefore, by Lemma~\ref{lemma:concavity}, the overall Hamiltonian 
is concave in $v_i$ for each fixed $t,x,p$ and $v_j$ with $j \neq i$. Since a non-constant continuous concave 
function over a compact set achieves its minimum on the boundary of the 
compact set (see~\cite{SCC.Zangwil1967}, Theorem~3), we can conclude that
\[
v \mapsto H(x^*(t), p^*(t), U_v, t)
\]
achieves its minimum at the boundary of the range of the embedded mode–signal domain, that is, the boundary of the set of all $b$ embedding variables.
At these boundary points, the embedded variables take binary values 
corresponding to feasible switching signals of the original SOCP. Hence, 
every optimal solution of the MEOCP corresponds to a feasible solution of 
the SOCP.
\end{proof}

The result demonstrates that the modification introduced in the MEOCP not only enables the use of continuous optimization techniques but also guarantees that the resulting solutions remain valid for the original SOCP.

\section{Simulation Results}
\label{sec:Simulation}
This section presents two simulation studies that demonstrate the effectiveness of the proposed MEOCP framework. The first case involves a nonlinear Three-Tanks System, used to compare the proposed method against the Mode Insertion Gradient (MIG) approach \cite{SCC.Wardi.Egerstedt.ea2014}. The second case considers a spacecraft rendezvous problem, which includes a quadratic terminal cost. The MIG method does not natively support terminal cost terms and requires additional effort to incorporate them. In contrast, the proposed MEOCP formulation allows such terms to be included directly and is easily implemented using standard optimal control solvers. 
Since in both case studies the running and terminal costs are continuously differentiable, the problems are compatible with collocation-based numerical methods, and we use GPOPS-II \cite{SCC.Patterson.Rao2014} to obtain the solutions.

\subsection{Three-Tanks System}
\begin{figure}[ht]
    \centering
    \includegraphics[width=0.6\linewidth]{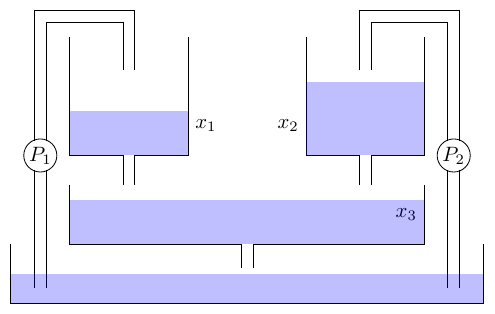}
    \caption{Schematics of the 3-tank system.}
    \label{Fig:3tanks}
\end{figure}

The schematics of the three-tank setup are shown in Fig.~\ref{Fig:3tanks}. The setup consists of three interconnected tanks with dynamics governed by
\begin{equation*}
\begin{bmatrix}
\dot{x}_1\\
\dot{x}_2\\
\dot{x}_3
\end{bmatrix}
=
\begin{bmatrix}
V_{p_1} - c_1\sqrt{x_1} \\
V_{p_2} - c_2\sqrt{x_2} \\
c_1\sqrt{x_1} + c_2\sqrt{x_2} - c_3\sqrt{x_3}
\end{bmatrix},
\end{equation*}
where $x_1, x_2, x_3$ denote the liquid levels in the tanks, $c_1 = c_2 = 1$ and $c_3 = 2$ are the outflow coefficients, and $V_{p_1}, V_{p_2} \in \{1, 2\}$ are the input flows determined by the active mode.

The control objective is to regulate the fluid levels toward the desired steady-state values  $x_1^d = x_2^d = 1$ and $x_3^d = 3$. To achieve this, we define the running cost as
\begin{equation*}
L(x) = d_1(x_3 - x_3^d)^2 + d_2(x_2 - x_1)^2,
\end{equation*}
where $d_1 = 3$ and $d_2 = 1$ are weighting coefficients. This cost penalizes both the deviation of the third tank from its target level and the imbalance between tanks 1 and 2.

To enforce bang-bang behavior in the embedded switching variables, we introduce the concave auxiliary cost term
\begin{equation*}
L_M(v) = \alpha \left[ v_0(1 - v_0) + v_1(1 - v_1) \right],
\end{equation*}
where $\alpha = 0.1$ is a design parameter. This term promotes binary values of the relaxed variables $v_0$ and $v_1$ by penalizing fractional solutions.

In the Three-Tanks System simulation, the proposed MEOCP method achieved a cost of $J^* = 31.5233$ with a simulation time of only 18.94 seconds. The simulations for both MEOCP and the Mode Insertion Gradient (MIG) method were performed on the same computational platform, featuring an Intel(R) Core(TM) i9-14900 CPU @ 2.00~GHz, 64.0~GB RAM, and MATLAB R2024a. Since MEOCP reformulates the switched optimal control problem as a continuous one, it enables the use of well-established numerical optimization tools developed for continuous systems. In this work, we employed \texttt{GPOPS-II}, a direct collocation-based optimal control solver, to efficiently solve the resulting embedded problem.

In contrast, the MIG method discretizes the time horizon using a fixed sample time $dt$ and performs switching evaluations at each discrete point. As shown in Table~\ref{Tab:MIG_Cost}, decreasing $dt$ from 0.01 to 0.001 results in a dramatic increase in simulation time across all tested values of $N$. For example, the simulation time grows from 10.64 seconds to 111.32 seconds for $N = 100$, and from 157.58 seconds to 2981.17 seconds for $N = 1000$. Although finer time steps result in slightly improved cost values, the rapidly increasing computational burden makes MIG less scalable and less practical for high-resolution applications.

\begin{figure}[ht]
    \centering

    \begin{subfigure}[t]{\linewidth}
        \centering
        \includegraphics[width=0.8\linewidth]{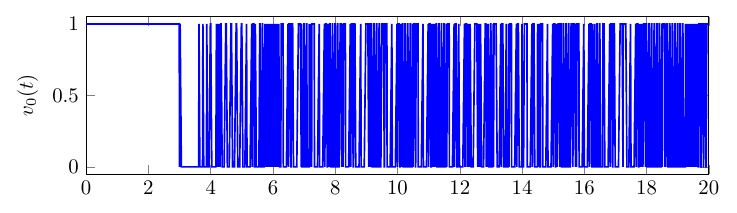}
    \end{subfigure}

    \begin{subfigure}[t]{\linewidth}
        \centering
        \includegraphics[width=0.8\linewidth]{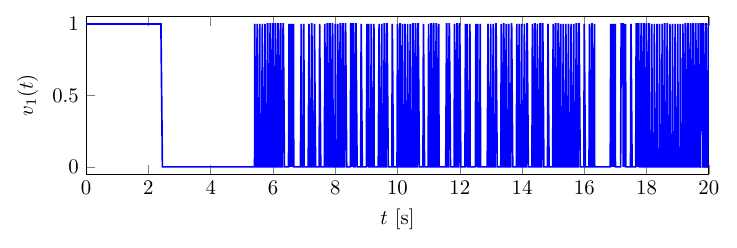}
    \end{subfigure}

    \caption{Time evolution of embedded binary switching variables $v_0(t)$ and $v_1(t)$ for the three-tanks MEOCP simulation.}
    \label{Fig:MEOCP_vi}
\end{figure}

\begin{figure}[ht]
    \centering
    \includegraphics[width=0.8\linewidth]{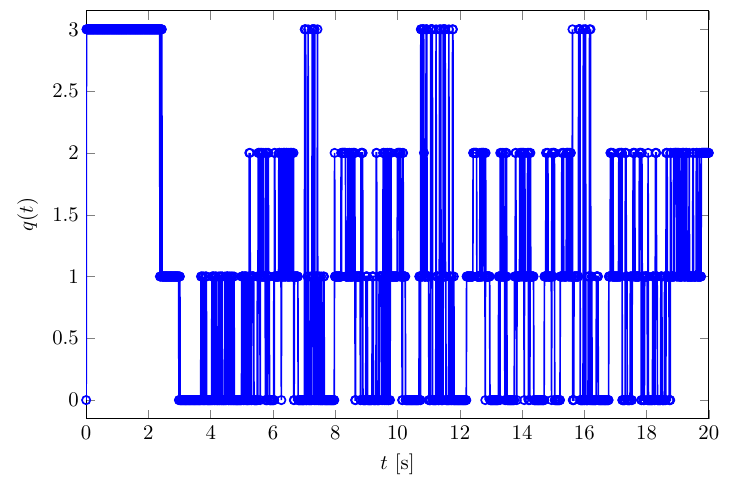}
    \caption{Mode index trajectory during the three-tanks simulation under the MEOCP solution.}
    \label{Fig:MEOCP_q}
\end{figure}

\begin{figure}[ht]
    \centering
    \includegraphics[width=0.85\linewidth]{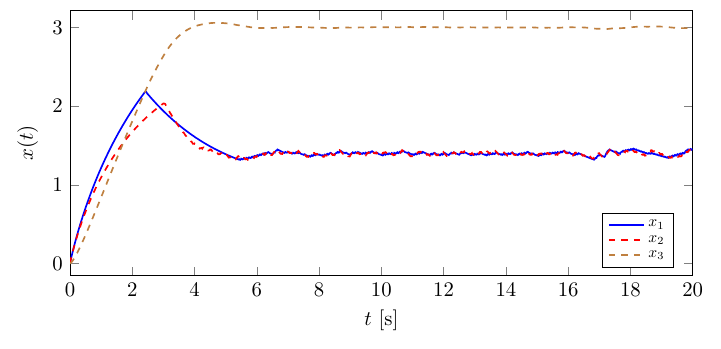}
    \caption{State trajectory of the three-tanks system under the MEOCP solution.}
    \label{Fig:MEOCP_solution}
\end{figure}

These results indicate that the MEOCP achieves competitive performance with significantly reduced computational effort. The binary switching variables $v_0(t), v_1(t) \in \{0,1\}$ are shown in Fig.~\ref{Fig:MEOCP_vi}, and the corresponding decoded switching signal $q(t) = (v_1(t), v_0(t))_2 \in \{0, 1, 2, 3\}$ is shown in Fig.~\ref{Fig:MEOCP_q} and  the resulting tank level trajectories are illustrated in Fig.~\ref{Fig:MEOCP_solution}.

\begin{table}
\centering
\caption{Mode Insertion Method Results: Cost $J^*$ and Simulation Time (ST)}
\resizebox{0.8\linewidth}{!}{%
\begin{tabular}{|c|c|c|}
\hline
\textbf{N} & \textbf{dt = 0.01} & \textbf{dt = 0.001} \\
\hline
100 & $J^*=32.1564$; ST=10.64 s & $J^*=32.1216$; ST=111.32 s \\
\hline
200  & $J^*=31.8507$; ST=23.28 s & $J^*=31.7329$; ST=244.38 s \\
\hline
300  & $J^*=31.7919$; ST=36.27 s & $J^*=31.6434$; ST=972.41 s \\
\hline
500  & $J^*=31.7738$; ST=68.32 s & $J^*=31.5788$; ST=677.50 s \\
\hline
1000 & $J^*=31.7738$; ST=157.58 s & $J^*=31.5461$; ST=2981.17 s \\
\hline
\end{tabular}
}
\label{Tab:MIG_Cost}
\end{table}

\subsection{Rendezvous problem}
\begin{figure}[htbp]
    \centering
    \includegraphics[width=0.6\linewidth]{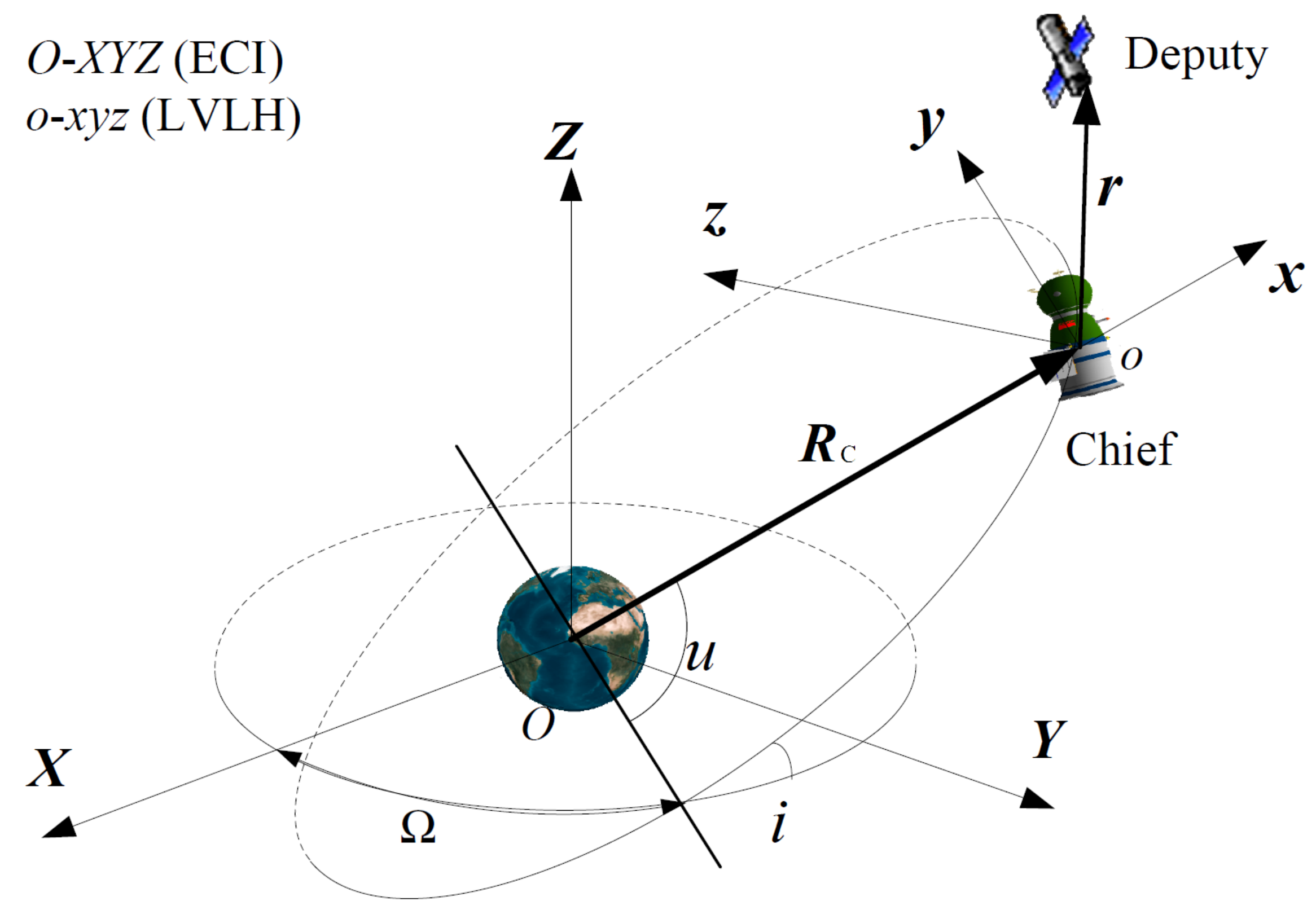}
    \caption{Schematic representation of the chief and deputy satellites in the Earth-centered inertial (ECI) and local-vertical local-horizontal (LVLH) frames.}
    \label{Fig.R.LVLH frame}
\end{figure}

For the second simulation, we consider a spacecraft rendezvous problem in which a thrust-actuated deputy satellite must intercept a chief satellite moving in a circular orbit. The equations of motion are derived by starting in the earth-centered inertial (ECI) frame, transforming to a rotating local-vertical local-horizontal (LVLH) frame attached to the chief, and then applying non-dimensionalization followed by simplification under the circular orbit assumption. A schematic illustrating the positions of the chief and deputy satellites in both the ECI and LVLH frames is shown in Fig.~\ref{Fig.R.LVLH frame}. Throughout this section, bold symbols denote physical vectors in $\mathbb{R}^3$, while their corresponding non-bold symbols denote magnitudes, defined as the Euclidean norm (e.g., $r = \lVert \mathbf{r} \rVert$).
\paragraph{Equations of motion in the ECI and LVLH frame}  
Let \( \mathbf{R}_c \) and \( \mathbf{R}_d \) denote the positions vector of the chief and deputy satellites, respectively, in the ECI frame. The gravitational equations of motion are  
\[
\ddot{\mathbf{R}}_c = -\mu \frac{\mathbf{R}_c}{R_c^3}, \qquad
\ddot{\mathbf{R}}_d = -\mu \frac{\mathbf{R}_d}{R_d^3} + \mathbf{u},
\]
where $\mu$ is Earth’s gravitational parameter and $\mathbf{u} = [u_x,\,u_y,\,u_z]^\top$ is the thrust acceleration vector of the deputy satellite. The relative position is defined as $\mathbf{r} = \mathbf{R}_d - \mathbf{R}_c$.

By applying the transport theorem for acceleration in a rotating frame, we relate the inertial dynamics to those observed in a rotating frame, resulting in the expression
\begin{align*}
\left( \frac{\mathrm{d}^2 \mathbf{r}}{dt^2} \right)_{\text{rot}} 
&= \left( \frac{\mathrm{d}^2 \mathbf{r}}{dt^2} \right)_{\text{inertial}} 
\ - \ \underbrace{2 \boldsymbol{\omega} \times \dot{\mathbf{r}}}_{\text{Coriolis force}}
\quad - \underbrace{\dot{\boldsymbol{\omega}} \times \mathbf{r}}_{\text{Euler force}} 
\ - \ \underbrace{\boldsymbol{\omega} \times (\boldsymbol{\omega} \times \mathbf{r})}_{\text{Centrifugal force}}
\end{align*}
which leads to the nonlinear relative dynamics
\begin{align*}
\ddot{x} - 2 \omega \dot{y} - \omega^2 x - \dot{\omega} y + \mu \left( \frac{R_c + x}{R_d^3} - \frac{1}{R_c^2} \right) &= u_x,  \\
\ddot{y} + 2 \omega \dot{x} - \omega^2 y + \dot{\omega} x + \mu \frac{y}{R_d^3} &= u_y,  \\
\ddot{z} + \mu \frac{z}{R_d^3} &= u_z, 
\end{align*}
where \( R_d = \sqrt{(R_c + x)^2 + y^2 + z^2} \).
\paragraph{Circular orbit simplification.}  
For this simulation, we assume the chief satellite follows a circular orbit of constant radius \( R \). In this case, the angular velocity \( \boldsymbol{\omega} \) remains constant in both magnitude and direction,
\[
\boldsymbol{\omega} = 
\begin{bmatrix}
0 \\ 0 \\ \omega
\end{bmatrix},
\quad
\omega = \sqrt{\mu / R^3},
\quad
\dot{\boldsymbol{\omega}} = 0.
\]
We introduce scaled variables using the reference length \( R \) and time constant \( \tau = 1/\omega \). The normalized position components are defined as  
\[
\bar{x} = \frac{x}{R}, \quad
\bar{y} = \frac{y}{R}, \quad
\bar{z} = \frac{z}{R},
\]
and the corresponding relative distance becomes  
\[
\bar{R}_d = \frac{R_d}{R} = \sqrt{(1 + \bar{x})^2 + \bar{y}^2 + \bar{z}^2},
\]
while the control input is scaled by  
\[
\bar{u}_i = \frac{u_i}{R \omega^2}, \quad i \in \{x, y, z\}.
\]
Under this transformation, the equations of motion reduce to
\begin{align*}
\ddot{\bar{x}} - 2\dot{\bar{y}} + (1 + \bar{x})\left( \frac{1}{\bar{R}_d^3} - 1 \right) &= \bar{u}_x,  \\
\ddot{\bar{y}} + 2\dot{\bar{x}} + \bar{y}\left( \frac{1}{\bar{R}_d^3} - 1 \right) &= \bar{u}_y, \\
\ddot{\bar{z}} + \bar{z} \left( \frac{1}{\bar{R}_d^3} \right) &= \bar{u}_z. 
\end{align*}
\paragraph{Planar dynamics and first-order form.}  
For simplicity, we restrict attention to planar motion in the \( \bar{x}\bar{y} \)-plane by assuming \( \bar{z} \equiv 0 \), so the vertical dynamics in Equation (14) are omitted. The state and control vectors are defined as  
\[
\xi = 
\begin{bmatrix}
\xi_1 \\ \xi_2 \\ \xi_3 \\ \xi_4
\end{bmatrix}
=
\begin{bmatrix}
\bar{x} \\ \bar{y} \\ \dot{\bar{x}} \\ \dot{\bar{y}}
\end{bmatrix},
\quad
\underline{u} = 
\begin{bmatrix}
u_1 \\ u_2
\end{bmatrix}
=
\begin{bmatrix}
\bar{u}_x \\ \bar{u}_y
\end{bmatrix},
\]
so the dynamics can be written as a first-order affine control system
\[
\dot{\xi} = f(\xi) + \underline{u},
\]
where the drift term is given by

e planar case, the relative distance reduces to \( \bar{R}_d = \sqrt{(1 + \xi_1)^2 + \xi_2^2} \),  
where a schematic of the orbital frame in the 
x-y plane is shown in Fig.~\ref{Fig:R2DOrbital}.

The control input is restricted to a finite set of five admissible thrust vectors, corresponding to either no thrust or unit thrust applied along a single axis direction (positive or negative 
x or y, with at most one thruster active at any time,
\[
\mathcal{U} = \left\{
\begin{bmatrix} 0 \\ 0 \end{bmatrix},
\begin{bmatrix} u_0 \\ 0 \end{bmatrix},
\begin{bmatrix} -u_0 \\ 0 \end{bmatrix},
\begin{bmatrix} 0 \\ u_0 \end{bmatrix},
\begin{bmatrix} 0 \\ -u_0 \end{bmatrix}
\right\},
\]
where \( u_0 > 0 \) is the fixed thrust magnitude in either direction.

The objective is to guide the deputy satellite to rendezvous with the chief while minimizing the quadratic cost
\[
J = \xi(t_f)^\top S \xi(t_f) + \int_{t_0}^{t_f} \xi(t)^\top Q \xi(t) \, \mathrm{d}t, 
\]
where \( Q \succ 0 \) is the state tracking matrix, and \( S \succ 0 \) is the terminal penalty matrix.

To implement the control logic using the MEOCP framework, we encode the five thrust modes using three binary variables \( v_0(t), v_1(t), v_2(t) \in [0,1] \), where the mode index is given by  
\[
q(t) = (v_2(t), v_1(t), v_0(t))_2 \in \{0,1,\dots,7\},
\]
and valid modes correspond to \( q \in \mathcal{Q} = \{0,1,2,3,4\} \).

To enforce bang-bang behavior and eliminate invalid configurations, we augment the cost functional as  
\[
J = \xi(t_f)^\top S \xi(t_f) + \int_{t_0}^{t_f} \left[ \xi(t)^\top Q \xi(t) + L_5(v(t)) \right] \mathrm{d}t, 
\]
where the penalty function is
\[
L_5(v) = \alpha \sum_{i=0}^{2} v_i(1 - v_i) + \beta \sum_{k=5}^{7} \prod_{i \,:\, k_i = 1} v_i,
\]
The first term discourages fractional values of $v_i$, while the second term penalizes binary combinations corresponding to invalid mode indices $q \in \{5,6,7\}$.

For simulation, we consider two satellites in coplanar circular orbits. The chief satellite follows a circular orbit of radius $R_c = 21000$ km, while the deputy satellite begins its maneuver from a lower circular orbit with initial radius $R_d = 18500$ km. Both satellites are assumed to be phase-aligned at the onset of the maneuver, meaning they share the same angular position at time $t = t_0$.

We normalize the system using the chief orbital radius $R = R_c = 21000$ km and select the corresponding orbital time constant $\tau = 4820$ s. The initial condition for the relative state in the non-dimensional LVLH frame is set as $\xi(0) = [-0.119,\ 0,\ 0,\ 0.065]^\top$.

The available control is the set of five discrete ON/OFF thrust modes, where the non-dimensional thrust magnitude is selected as $u_0 = 3$, which corresponds to a dimensional acceleration of $a = R\omega^2 \cdot u_0 \approx 0.9039$ m/s$^2$.

The cost weight matrices are selected as $Q = \mathrm{diag}(100, \allowbreak  100, 1, 1)$ and $S = 10Q$ to emphasize terminal accuracy in the position coordinates. The set of admissible states  is restricted to the bounded domain
\[
\Omega = \left\{ \xi \in \mathbb{R}^4 \,\middle|\, -0.35 \leq \xi_i \leq 0.35,\ i = 1,2,3,4 \right\}
\]
to ensure that the states remains within physically meaningful limits during the maneuver.

\begin{figure}[ht]
    \centering

    \begin{subfigure}[t]{\linewidth}
        \centering
        \includegraphics[width=0.8\linewidth]{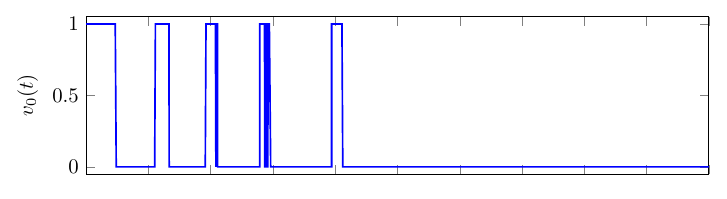}
    \end{subfigure}

    \begin{subfigure}[t]{\linewidth}
        \centering
        \includegraphics[width=0.8\linewidth]{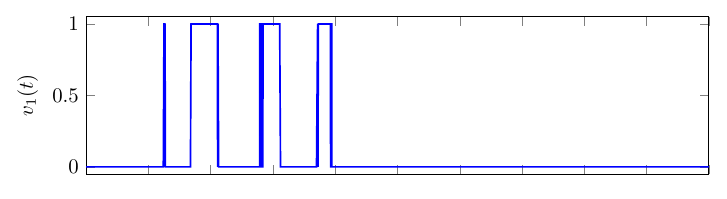}
    \end{subfigure}

    \begin{subfigure}[t]{\linewidth}
        \centering
        \includegraphics[width=0.8\linewidth]{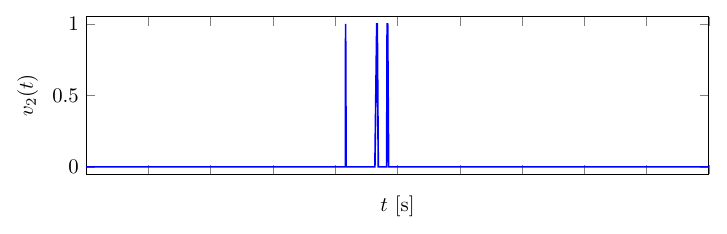}
    \end{subfigure}

    \caption{Time evolution of embedded binary variables $v_0(t)$, $v_1(t)$, and $v_2(t)$ for the rendezvous MEOCP simulation.}
    \label{Fig:RMEOCP_vi}
\end{figure}

\begin{figure}[ht]
    \centering
    \includegraphics[width=0.8\linewidth]{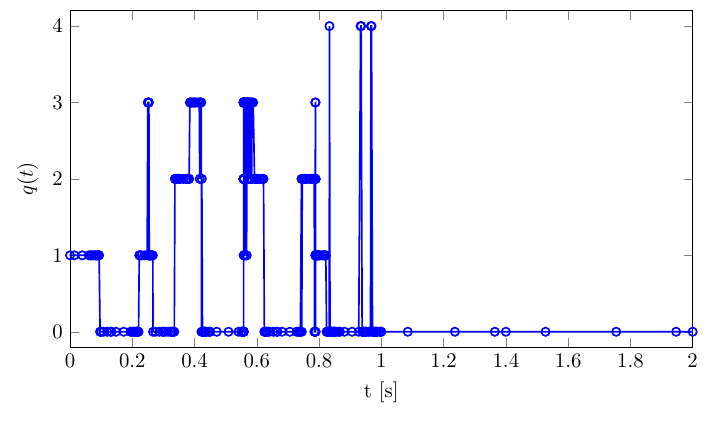}
    \caption{Mode index trajectory during the rendezvous maneuver under the MEOCP solution.}
    \label{Fig:RMEOCP_q}
\end{figure}

\begin{figure}[ht]
    \centering
    \includegraphics[width=0.85\linewidth]{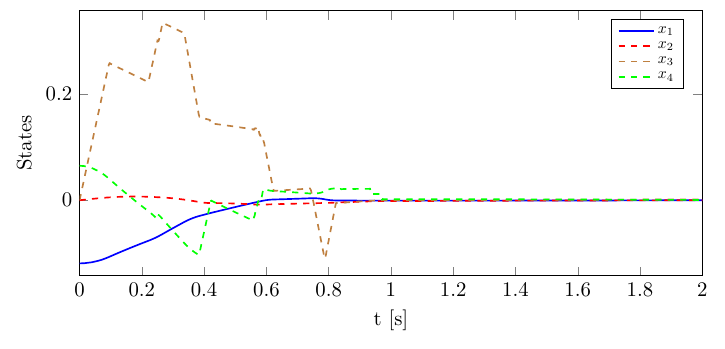}
    \caption{State trajectory in the LVLH frame during the rendezvous maneuver under the MEOCP solution.}
    \label{Fig:RMEOCP_LVLH}
\end{figure}

\begin{figure}[ht]
    \centering
    \includegraphics[width=0.6\linewidth]{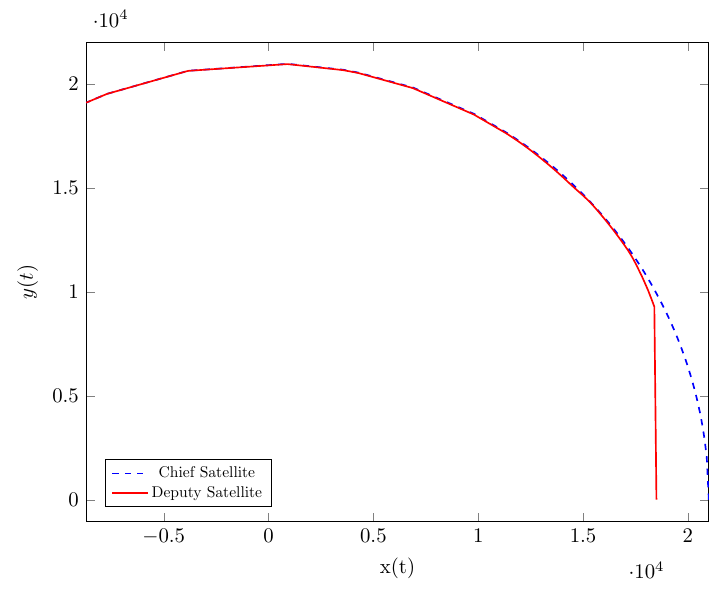}
    \caption{Trajectory of the deputy satellite in the Earth-centered inertial (ECI) frame during the rendezvous maneuver.}
    \label{Fig:RMEOCP_ECI}
\end{figure}

Fig. \ref{Fig:RMEOCP_vi} illustrates the time evolution of the embedded binary switching variables $ v_0(t), v_1(t), v_2(t) \in [0,1] $. The corresponding decoded mode index $ q = (v_2, v_1, v_0)_2 \in \{0,1,2,3,4\} $, shown in Fig. \ref{Fig:RMEOCP_q}, reveals the sequence of applied thrust inputs throughout the maneuver.
Fig. \ref{Fig:RMEOCP_LVLH} presents the state trajectory  $x$  in the scaled LVLH frame, including both position and velocity components. The successful convergence to the target state, as demonstrated in Fig. \ref{Fig:RMEOCP_LVLH} confirms the effectiveness of the MEOCP-generated control policy.
Finally, the trajectory of the deputy spacecraft in the Earth-centered inertial (ECI) frame, projected onto the orbital \( x\text{-}y \) plane, is shown in Fig.~\ref{Fig:RMEOCP_ECI}. This plot illustrates the gradual approach of the deputy toward the chief along a dynamically feasible transfer path.
\begin{figure}[ht]
    \centering
    \includegraphics[width=0.35\linewidth]{Figures/Rendezvous/2D_Orbital.pdf}
    \caption{Orbital frame schematic in the $x$--$y$ plane showing the deputy and chief satellite positions relative to the center of gravity.}
    \label{Fig:R2DOrbital}
\end{figure}

\section{Discussion}
The numerical simulations confirm that the proposed MEOCP framework effectively generates feasible, bang-bang switching strategies for multi-mode systems while maintaining computational tractability. In both test cases—the nonlinear three-tanks system and the spacecraft rendezvous problem—the embedded formulation enabled the use of conventional continuous optimal control solvers, such as GPOPS-II, without requiring problem-specific modifications.

Compared to the mode insertion gradient (MIG) method, which relies on fixed time discretization and iterative updates to the switching sequence, the MEOCP approach reformulates the original problem into a continuous optimal control framework. This eliminates the need for explicitly tracking switching events and sequences, thereby avoiding the combinatorial complexity inherent in discrete-time approaches. As demonstrated in the three-tanks example, MIG can reach similar minimum cost values as MEOCP by using small time steps, at the expense of  significantly increased  overall simulation time and computational burden.

A notable advantage of MEOCP is its flexibility in incorporating terminal cost terms and state constraints. Because the formulation is continuous, such constraints can be directly encoded in the problem definition and handled natively by numerical solvers such as GPOPS-II. This enables seamless inclusion of endpoint objectives and path constraints without requiring algorithmic modifications or custom extensions to the optimization routine. In contrast, adapting the MIG framework to accommodate these elements typically demands nontrivial modifications to its formulation, which are not addressed in ~\cite{SCC.Wardi.Egerstedt.ea2014}. This flexibility enhances the applicability of MEOCP in complex and constrained optimal control settings.

In summary, the proposed MEOCP framework combines scalability, solver compatibility, and feasibility enforcement in a unified formulation. It offers a practical and efficient alternative to mode insertion-based methods, particularly for systems with many discrete modes, complex constraints, and terminal objectives.
\section{Conclusion}
This paper presents an embedding-based approach for solving switched optimal control problems (SOCPs) with an arbitrary number of subsystems. The discrete switching signal, which represents the activation of different modes, is initially encoded using binary variables. These binary variables are then replaced with continuous embedded variables that can take intermediate values between zero and one, resulting in the formulation of an embedded optimal control problem (EOCP). A novel approach for designing a concave function in the proper dimension is presented, which, when added to the main cost function, yields the corresponding modified embedded optimal control problem (MEOCP). This ensures the system attains a bang-bang solution, thereby providing the control strategy as a solution to the original SOCP.

The developed method is compared to the mode insertion method, a widely recognized approach for addressing multi-mode switching systems. Unlike the mode insertion method, the proposed embedding-based approach allows for the direct application of conventional techniques developed for continuous optimal control problems. Moreover, it simplifies the incorporation of state constraints and end cost terms, which are challenging to handle in the mode insertion framework. This flexibility and ease of handling constraints, combined with the ability to directly derive bang-bang solutions, make the proposed method a powerful and practical alternative for solving complex multi-mode SOCPs.

Despite these advantages, the proposed method has its limitations. Specifically, it currently does not address dwell-time constraints, which are crucial in many real-world applications to ensure stability and practicality in switching systems.
Future work will focus on extending the framework to incorporate such constraints, thereby allowing the method to address a wider range of scenarios.


\section{Appendix}

\setcounter{lemma}{0} 
\renewcommand{\thelemma}{\ref{lemma:concavity}} 

\begin{lemma}
If $f:\mathbb{R}^b \to \mathbb{R}$ is an affine function, $g:\mathbb{R}^b \to \mathbb{R}$ a concave function, and  $h:\mathbb{R}^b \to \mathbb{R}$ is defined as $h(v) := f(v) + g(v)$, then $h$ is concave.
\end{lemma}


\begin{proof}
Let $f: \mathbb{R}^b \rightarrow \mathbb{R}$ be an affine function and $g: \mathbb{R}^b \rightarrow \mathbb{R}$ a concave function. Let $v^{(1)}, v^{(2)} \in \mathbb{R}^b$ and let $\lambda \in [0,1]$. Define the convex combination
\begin{equation*}
v := \lambda v^{(1)} + (1 - \lambda)v^{(2)}.
\end{equation*}

Define $h: \mathbb{R}^b \rightarrow \mathbb{R}$ by
\begin{equation*}
h(v) := f(v) + g(v).
\end{equation*}

Since $f$ is affine, there exist $a \in \mathbb{R}^b$ and $c \in \mathbb{R}$ such that
\begin{equation*}
f(v) = a^\top v + c.
\end{equation*}
Since $g$ is concave, it satisfies
\begin{equation*}
g(v) \geq \lambda g(v^{(1)}) + (1 - \lambda) g(v^{(2)}).
\end{equation*}
We calculate $h(v)$ as follows.
\begin{align*}
h(v) &= f(v) + g(v) \nonumber \\
&= a^\top v + c + g(v) \nonumber \\
&\geq a^\top (\lambda v^{(1)} + (1 - \lambda)v^{(2)}) + c + \lambda g(v^{(1)}) \nonumber \\
& + (1 - \lambda) g(v^{(2)}) \nonumber \\
&= \lambda (   f(v^{(1)})    + g(v^{(1)})  )+(1 - \lambda)(    f(v^{(2)})    + g(v^{(2)})  )\nonumber \\
&= \lambda h(v^{(1)}) + (1 - \lambda) h(v^{(2)}).
\end{align*}

Therefore, $h(v) \geq \lambda h(v^{(1)}) + (1 - \lambda) h(v^{(2)})$, which proves that $h$ is concave on $\mathbb{R}^b$.
\end{proof}

 

\end{document}